\documentclass[aip,jcp,preprint,groupedaddress]{revtex4-2}%
\usepackage{graphicx}
\usepackage{tabularx}
\usepackage{dcolumn}
\usepackage{longtable}
\usepackage{tensor}
\usepackage{placeins}
\usepackage{bm}
\usepackage{amsmath}
\usepackage{amsfonts}
\usepackage{amssymb}
\usepackage{float}
\usepackage{xcolor}%
\providecommand{\U}[1]{\protect\rule{.1in}{.1in}}
\providecommand{\U}[1]{\protect\rule{.1in}{.1in}}
\begin{document}
\title{On-the-fly \textit{ab initio} semiclassical evaluation of third-order response
functions for two-dimensional electronic spectroscopy}
\author{Tomislav Begu\v{s}i\'{c}}
\email{tomislav.begusic@epfl.ch}
\author{Ji\v{r}\'i Van\'i\v{c}ek}
\email{jiri.vanicek@epfl.ch}
\affiliation{Laboratory of Theoretical Physical Chemistry, Institut des Sciences et
Ing\'enierie Chimiques, Ecole Polytechnique F\'ed\'erale de Lausanne (EPFL),
CH-1015, Lausanne, Switzerland}
\date{\today}

\begin{abstract}
Ab initio computation of two-dimensional electronic spectra is an
expanding field, whose goal is improving upon simple, few-dimensional
models often employed to explain experiments. Here, we propose an accurate and
computationally affordable approach, based on the single-trajectory
semiclassical thawed Gaussian approximation, to evaluate two-dimensional
electronic spectra. Importantly, the method is exact for arbitrary harmonic
potentials with mode displacement, changes in the mode frequencies, and
inter-mode coupling (Duschinsky effect), but can also account partially for
the anharmonicity of the involved potential energy surfaces. We test its
accuracy on a set of model Morse potentials and use it to study anharmonicity
and Duschinsky effects on the linear and two-dimensional electronic spectra of
phenol. We find that in this molecule, the anharmonicity effects are weak, whereas the Duschinsky rotation and the changes in the
mode frequencies must be included in accurate simulations. In contrast, the widely used
displaced harmonic oscillator model captures only the basic physics of the
problem but fails to reproduce the correct vibronic lineshape.

\end{abstract}
\maketitle

\graphicspath{{"C:/Users/GROUP LCPT/Documents/Group/Tomislav/ThirdOrder/figures/"}
{./figures/}{C:/Users/Jiri/Dropbox/Papers/Chemistry_papers/2020/ThirdOrder/figures/}}

\section{\label{sec:intro}Introduction}

Electronic spectroscopy allows us to study excited electronic states and
light-induced nuclear dynamics. To track this ultrafast dynamics on
femtosecond time scales, a range of time-resolved and two-dimensional
spectroscopic techniques were developed. The complex signals obtained in these
experiments are, however, difficult to interpret without the help of
theoretical modeling.\cite{Conti_Garavelli:2020}

Most models for describing two-dimensional electronic spectra treat the
electronic states as the system and include nuclear dynamics only approximately, as bath effects.
The simplest approach assumes that nuclear degrees of freedom
induce a Gaussian-like or Lorentzian-like broadening, neglecting completely
the coherent nuclear dynamics. Such models are appropriate only if the
coherent dynamics is strongly suppressed by the surrounding solvent dynamics.
To account for inhomogeneous (static) broadening, the energy gap between the
electronic states can be averaged over snapshots of different arrangements of
the
environment.\cite{Rivalta_Garavelli:2014,Giussani_Nenov:2017,Borrego-Varillas_Cerullo:2019}
Alternatively, a swarm of trajectories could be used in Kubo-type
calculations, where each trajectory is equipped with a phase, obtained
from the time integral of the energy gap between the involved electronic
states along the trajectory, and the correlation functions are averaged over
the full
ensemble;\cite{Mukamel:1982,Shemetulskis_Loring:1992,Li_Fang:1996,VanDerVegte_Jansen:2013,Tempelaar_Jansen:2013,Petit_Subotnik:2014}
these approaches are also known as phase averaging,\cite{book_Mukamel:1999}
Wigner-averaged classical
limit,\cite{Egorov_Rabani:1998,Egorov_Rabani:1999,Shi_Geva:2004} or dephasing
representation.\cite{Vanicek:2004, Vanicek:2006,Zimmermann_Vanicek:2014} Such
methods are accurate when the curvatures of different potential energy
surfaces are similar and in the limit of strong dephasing, i.e., for short
times. Even in this case, however, if the on-the-fly dynamics is
performed with ab initio electronic structure methods, evolving the full
ensemble of classical trajectories can quickly become prohibitively expensive.
To account for both intramolecular and intermolecular nuclear dynamics in
two-dimensional spectroscopy, one can use the multimode Brownian oscillator
model,\cite{book_Mukamel:1999,Nemeth_Sperling:2008,Schlau-Cohen_Fleming:2011,Caram_Engel:2012}
which considers a few primary harmonic modes coupled to a large number of
low-frequency bath oscillators. For this model, the spectra can be computed
analytically; moreover, the parameters of the model can be computed
efficiently from a single ab initio classical trajectory, as demonstrated in
Refs.~\onlinecite{Nenov_Garavelli:2015, Picchiotti_Garavelli:2019}. This
allows one to perform electronic structure computations at a high level of
theory, using, for example, post-Hartree--Fock multiconfigurational
wavefunction methods. Yet, the approach is limited to modeling the molecule as
a set of few uncoupled displaced harmonic oscillators. Such a simple
description is inadequate in systems that exhibit strong mode-mode coupling,
changes in the force constants between the ground and excited electronic
states, or anharmonicity effects. A generalization to a set of uncoupled
harmonic oscillators with both displacement and a possible change in the force
constant was proposed by Fidler and Engel, who used the approximate
third-order cumulant expansion.\cite{Fidler_Engel:2013} Very recently, the third-order cumulant expansion was also studied as an
approximate way to account for the mode-mode coupling (Duschinsky effect) in
both linear\cite{Zuehlsdorff_Isborn:2019} and
nonlinear\cite{Zuehlsdorff_Isborn:2020} spectra.

There has been little development in the ab initio simulation of
two-dimensional electronic spectra beyond the standard semiclassical methods
or displaced harmonic models. To account for anharmonicity
effects\cite{Pour_Hauer:2017,Anda_Hansen:2018} or more general coupled
oscillators, one is forced to employ computationally expensive exact quantum
dynamics
methods,\cite{Schubert_Engel:2011,Krcmar_Domcke:2015,Sala_Egorova:2016,
Choi_Vanicek:2019,Roulet_Vanicek:2019,Choi_Vanicek:2019a} such as different
flavors of the multiconfigurational time-dependent Hartree (MCTDH)
method\cite{book_MCTDH,Krcmar_Domcke:2013}, or the hierarchical equations of
motion.\cite{Xu_Yan:2011,Tanimura:2020} These methods require the
pre-computation of the full potential energy surfaces and are not suitable for
a first-principles on-the-fly implementation. First-principles
multi-trajectory semiclassical
approaches\cite{Tatchen_Pollak:2009,Ceotto_Atahan:2009,Ceotto_Atahan:2009a,Buchholz_Ceotto:2016,Buchholz_Ceotto:2017,Gabas_Ceotto:2017,Gabas_Ceotto:2018,Bonfanti_Pollak:2018,Micciarelli_Ceotto:2019}
and direct quantum dynamics methods, which often use multiple
Gaussians,\cite{Martinez_Levine:1996a,Curchod_Martinez:2018,Makhov_Shalashilin:2017,Sulc_Vanicek:2013,Worth_Burghardt:2004,Richings_Lasorne:2015,Bonfanti_Pollak:2018,Polyak_Knowles:2019}
also called coherent or Davydov
states,\cite{Sun_Zhao:2015,Zhou_Zhao:2016,Werther_Grossmann:2020} to represent
the evolving wavepacket, are impractical due to the large number of required
ab initio evaluations.

Here, we propose an efficient semiclassical method to evaluate vibrationally
resolved two-dimensional electronic spectra. The approach, based on Heller's
single-trajectory thawed Gaussian approximation,\cite{Heller:1975} accounts
for inter-mode coupling, changes in the force constants, and, at least
partially, for the anharmonicities of the ground- and excited-state potential
energy surfaces. First, we study how the accuracy of the method depends on the
degree of anharmonicity in the one-dimensional Morse system. The results are
compared with the exact benchmark and with the harmonic approximation, which
neglects the anharmonicity completely. Second, we analyze the effects of
Duschinsky coupling and anharmonicity on the linear absorption and
two-dimensional spectra of phenol.

\section{\label{sec:theory}Theory}

\subsection{\label{subsec:polarization}Third-order response function}

The central object in all types of third-order electronic spectroscopy is the
third-order polarization\cite{book_Mukamel:1999,Gelin_Domcke:2009}
\begin{equation}
P^{(3)}(t)=\int_{0}^{\infty}dt_{3}\int_{0}^{\infty}dt_{2}\int_{0}^{\infty
}dt_{1}R(t_{3},t_{2},t_{1})E(t-t_{3})E(t-t_{3}-t_{2})E(t-t_{3}-t_{2}-t_{1}),
\label{eq:P_t}%
\end{equation}
where $E(t)$ is the electric field of light (without the polarization vector)
and
\begin{equation}
R(t_{3},t_{2},t_{1})=\left(  \frac{i}{\hbar}\right)  ^{3}\sum_{\alpha=1}%
^{4}[R_{\alpha}(t_{3},t_{2},t_{1})-R_{\alpha}(t_{3},t_{2},t_{1})^{\ast}]
\label{eq:R}%
\end{equation}
is the third-order response function, expressed in terms of correlation
functions
\begin{align}
R_{1}(t_{3},t_{2},t_{1})  &  =C(t_{2},t_{3},t_{1}+t_{2}+t_{3}),\label{eq:R1}\\
R_{2}(t_{3},t_{2},t_{1})  &  =C(t_{1}+t_{2},t_{3},t_{2}+t_{3}),\label{eq:R2}\\
R_{3}(t_{3},t_{2},t_{1})  &  =C(t_{1},t_{2}+t_{3},t_{3}),\label{eq:R3}\\
R_{4}(t_{3},t_{2},t_{1})  &  =C(-t_{3},-t_{2},t_{1}), \label{eq:R4}%
\end{align}
and
\begin{equation}
C(\tau_{a},\tau_{b},\tau_{c})=\text{Tr}[\hat{\rho}\hat{\mu}e^{i\hat{H}_{2}%
\tau_{a}/\hbar}\hat{\mu}e^{i\hat{H}_{1}\tau_{b}/\hbar}\hat{\mu}e^{-i\hat
{H}_{2}\tau_{c}/\hbar}\hat{\mu}e^{-i\hat{H}_{1}(\tau_{a}+\tau_{b}-\tau
_{c})/\hbar}]. \label{eq:C_tau}%
\end{equation}
In Eq.~(\ref{eq:C_tau}), $\hat{H}_{i}$ are the vibrational Hamiltonians of the
ground (\textquotedblleft$1$\textquotedblright) and excited (\textquotedblleft%
$2$\textquotedblright) electronic states, $\hat{\rho}=\exp(-\beta\hat{H}%
_{1})/\text{Tr}[\exp(-\beta\hat{H}_{1})]$ is the vibrational density operator
in the ground electronic state, and $\hat{\mu} = \hat{\mu}_{21}=\hat{\vec{\mu}}_{21}\cdot\vec{\epsilon}$ is the electronic transition
dipole moment projected on the polarization unit vector $\vec{\epsilon}$ of the external
electric field. Equations~(\ref{eq:R1})--(\ref{eq:R4}) rely on the following
assumptions: (i) due to the large gap between the electronic states, only the
ground electronic state is initially populated; (ii) Born--Oppenheimer
approximation, i.e., there is no population transfer under field-free
evolution; (iii) light pulses are linearly polarized in the same
direction $\vec{\epsilon}$; (iv) only two electronic states are involved. In
the following, we discuss how to evaluate the components $R_{\alpha}$ of the
response function (\ref{eq:R}).

\subsection{\label{subsec:zero_temp}Zero-temperature limit: Wavepacket
picture}

\begin{figure}
\includegraphics[width=\textwidth]{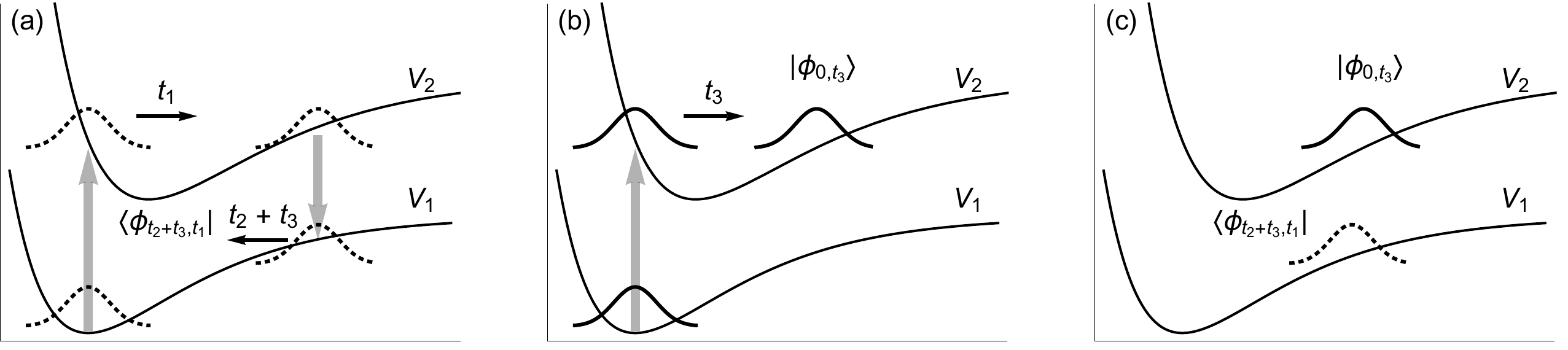} \caption{Evolution of the bra (a, dotted line) and ket (b, solid line) wavepackets of Eqs.~(\ref{eq:C_tau}) and (\ref{eq:C_wp}) for $\tau_a = t_1$, $\tau_b = t_2 + t_3$, and $\tau_c = t_3$. Their overlap (c) is the $R_3(t_3, t_2, t_1)$ component of the third-order response function $R(t_3, t_2, t_1)$ [see Eqs.~(\ref{eq:R3}), (\ref{eq:C_tau}), and (\ref{eq:C_wp})].\label{fig:R3_Scheme}}
\end{figure}

In the zero-temperature limit, we assume that only the ground
(\textquotedblleft$g$\textquotedblright) vibrational state $|1,g\rangle$ of
the ground electronic state is populated initially, i.e., $\hat{\rho
}=|1,g\rangle\langle1,g|$. Then, we may rewrite Eq.~(\ref{eq:C_tau}) in terms
of nuclear wavepackets:
\begin{align}
C(\tau_{a},\tau_{b},\tau_{c})  &  =\langle1,g|\hat{\mu}e^{i\hat{H}_{2}\tau
_{a}/\hbar}\hat{\mu}e^{i\hat{H}_{1}\tau_{b}/\hbar}\hat{\mu}e^{-i\hat{H}%
_{2}\tau_{c}/\hbar}\hat{\mu}e^{-i\hat{H}_{1}(\tau_{a}+\tau_{b}-\tau_{c}%
)/\hbar}|1,g\rangle\\
&  =\langle1,g|\hat{\mu}e^{i\hat{H}_{2}\tau_{a}/\hbar}\hat{\mu}e^{i\hat{H}%
_{1}\tau_{b}/\hbar}\hat{\mu}e^{-i\hat{H}_{2}\tau_{c}/\hbar}\hat{\mu
}|1,g\rangle e^{-i\omega_{1,g}(\tau_{a}+\tau_{b}-\tau_{c})}\\
&  =\langle1,g|\hat{\mu}e^{i\hat{H}_{2}^{\prime}\tau_{a}/\hbar}\hat{\mu
}e^{i\hat{H}_{1}^{\prime}\tau_{b}/\hbar}\hat{\mu}e^{-i\hat{H}_{2}^{\prime}%
\tau_{c}/\hbar}\hat{\mu}|1,g\rangle\\
&  =\langle\phi_{\tau_{b},\tau_{a}}|\phi_{0,\tau_{c}}\rangle, \label{eq:C_wp}%
\end{align}
where $\hbar\omega_{1,g}=\langle1,g|\hat{H}_{1}|1,g\rangle$, $\hat{H}%
_{i}^{\prime}=\hat{H}_{i}-\hbar\omega_{1,g}$, and
\[
\phi_{\tau,t}=e^{-i\hat{H}_{1}^{\prime}\tau/\hbar}\hat{\mu}e^{-i\hat{H}%
_{2}^{\prime}t/\hbar}\hat{\mu}|1,g\rangle.
\]
The result (\ref{eq:C_wp}) has an appealing interpretation in terms of bra and
ket wavepackets, which we represent pictorially for $R_{3}$ [Eq.~(\ref{eq:R3}%
)] in Fig.~\ref{fig:R3_Scheme}. The bra wavepacket is first evolved in the
excited electronic state for a time $\tau_{a}=t_{1}$ and then for a time
$\tau_{b}=t_{2}+t_{3}$ in the ground state, where it is a non-stationary
wavepacket due to the initial $t_{1}$ dynamics on the excited-state
potential energy surface. The ket wavepacket \textquotedblleft
waits\textquotedblright\ during the $t_{1}$ and $t_{2}$ times and is only
evolved for a time $\tau_{c}=t_{3}$ in the excited-state. This simple picture
has been discussed in the literature in the context of
pump-probe\cite{Pollard_Mathies:1990a} and
two-dimensional\cite{Schubert_Engel:2011} spectroscopy. In general,
during the $t_{1}$ (coherence) and $t_{3}$ (detection) times, the bra and ket
wavepackets evolve on different potential energy surfaces, i.e., the system is
in a state of electronic coherence; during $t_{2}$, also called population or
waiting time, both nuclear wavepackets are in the same electronic state, i.e.,
the system is in an electronic population
state.\cite{Schlau-Cohen_Fleming:2011}

The evaluation of $R_{\alpha}$ functions requires only one excited-state
wavepacket evolution up to time $t_{1} + t_{2} + t_{3}$ and, in addition,
wavepackets propagated in the ground electronic state starting from the
snapshots along the excited-state trajectory. Since such calculations would be
difficult to perform with multiple-trajectory direct dynamics methods, we
employ the efficient, single-trajectory thawed Gaussian approximation.

\subsection{\label{subsec:tga}Thawed Gaussian approximation}

Within the thawed Gaussian approximation, the time-dependent wavepacket takes
the form of a Gaussian
\begin{equation}
\psi_{t}(q)=e^{\frac{i}{\hbar}[\frac{1}{2}(q-q_{t})^{T}\cdot A_{t}%
\cdot(q-q_{t})+p_{t}^{T}\cdot(q-q_{t})+\gamma_{t}]}, \label{eq:gwp}%
\end{equation}
where $q_{t}$ and $p_{t}$ are $D$-dimensional position and momentum vectors,
$A_{t}$ is a complex and symmetric $D\times D$ matrix with positive-definite
imaginary part, and $\gamma_{t}$ is a complex number whose imaginary part
ensures the normalization. $D$ is the number of degrees of freedom. The
wavepacket (\ref{eq:gwp}) solves exactly the time-dependent Schr\"{o}dinger
equation
\begin{equation}
i\hbar|\dot{\psi}_{t}\rangle=[T(\hat{p})+V_{\text{LHA}}(\hat{q})]|\psi
_{t}\rangle,
\end{equation}
where $T(p)=\frac{1}{2}p^{T}\cdot m^{-1}\cdot p$, $m$ is a $D\times D$
symmetric mass matrix and
\begin{equation}
V_{\text{LHA}}(q)=V(q_{t})+V^{\prime}(q_{t})^{T}\cdot(q-q_{t})+\frac{1}%
{2}(q-q_{t})^{T}\cdot V^{\prime\prime}(q_{t})\cdot(q-q_{t}), \label{eq:lha}%
\end{equation}
is the local harmonic approximation of the true potential $V(q)$ about $q_{t}%
$, if the Gaussian's parameters satisfy the system\cite{Heller:1975}
\begin{align}
\dot{q}_{t}  &  =m^{-1}\cdot p_{t},\label{eq:q_t_dot}\\
\dot{p}_{t}  &  =-V^{\prime}(q_{t}),\label{eq:p_t_dot}\\
\dot{A}_{t}  &  =-A_{t}\cdot m^{-1}\cdot A_{t}-V^{\prime\prime}(q_{t}%
),\label{eq:A_t_dot}\\
\dot{\gamma}_{t}  &  =L_{t}+\frac{i\hbar}{2}\text{Tr}(m^{-1}\cdot A_{t}).
\label{eq:gamma_t_dot}%
\end{align}
In Eq.~(\ref{eq:gamma_t_dot}), $L_{t}=T(p_{t})-V(q_{t})$ is the Lagrangian of
the classical trajectory $(q_{t},p_{t})$. The above equations are interpreted
as follows: the phase-space center $(q_{t},p_{t})$ of the Gaussian
(\ref{eq:gwp}) evolves classically with the exact classical Hamiltonian, the
complex matrix $A_{t}$ evolves according to the Hessian computed at the
current $q_{t}$, and the complex number $\gamma_{t}$ is updated according to
the Lagrangian of the classical trajectory $(q_{t},p_{t})$ and the matrix
$A_{t}$. Since the only source of error is the local harmonic approximation
(\ref{eq:lha}), the thawed Gaussian propagation is exact for arbitrary,
multi-dimensional harmonic potentials.

The method was originally proposed for problems involving short-time
dynamics, such as photodissociation spectra.\cite{Lee_Heller:1982} However,
its accuracy appears to be surprisingly satisfactory in molecular systems even
at longer times, because many molecules are only weakly to moderately
anharmonic.\cite{Patoz_Vanicek:2018,Begusic_Vanicek:2018} Using a single
thawed Gaussian wavepacket, which is the essence of Heller's thawed Gaussian
approximation, is rather restrictive but also very efficient for on-the-fly
dynamics coupled to ab initio electronic structure. The on-the-fly ab initio
thawed Gaussian approximation\cite{Vanicek_Begusic:2021} proved useful in
treating efficiently anharmonicity effects on linear
absorption,\cite{Wehrle_Vanicek:2015,Patoz_Vanicek:2018,Begusic_Vanicek:2018,Prlj_Vanicek:2020}
emission,\cite{Wehrle_Vanicek:2014,Prlj_Vanicek:2020} and photoelectron
spectra,\cite{Wehrle_Vanicek:2015} as well as in understanding nuclei-induced
electronic decoherence in attosecond experiments.\cite{Golubev_Vanicek:2020}
Recently, we extended our on-the-fly implementation of the single-Gaussian
approach to simulate frequency- and time-resolved pump-probe
spectra,\cite{Begusic_Vanicek:2018a} similar to the earlier work by Rohrdanz and Cina\cite{Rohrdanz_Cina:2006} and Braun et al.\cite{Braun_Engel:1998} on model
potentials. Although ensembles of thawed Gaussians were largely
discarded and replaced by frozen Gaussians due to the numerical instabilities
that often appear in nonadiabatic dynamics simulations,
thawed Gaussians are being reintroduced, e.g., in the semiclassical
hybrid
dynamics\cite{Grossmann:2006,Goletz_Grossmann:2009,Buchholz_Jungwirth:2012} or
Gaussian-based MCTDH,\cite{Romer_Burghardt:2013,Eisenbrandt_Burghardt:2018,
Eisenbrandt_Burghardt:2018a} and especially for spectroscopic
applications,\cite{Kovac_Cina:2017,Picconi_Burghardt:2019,
Picconi_Burghardt:2019a, Picconi_Burghardt:2019b} due to their ability to
describe couplings between different degrees of freedom.\cite{Cheng_Cina:2014}

\subsection{\label{subsec:2DES}Two-dimensional electronic spectroscopy}

A variety of different third-order experiments can be simulated through the
computation of the response function (\ref{eq:R}).\cite{book_Mukamel:1999} For
example, the full response function is needed for evaluating transient
absorption spectra with finite-duration pulses.\cite{Pollard_Mathies:1992}
Here, we focus on the two-dimensional electronic spectra
\begin{equation}
S_{\alpha}(\omega_{3},\omega_{1})=\text{Re}\int_{0}^{\infty}dt_{3}\int%
_{0}^{\infty}dt_{1}R_{\alpha}(t_{3},0,t_{1})e^{i\omega_{3}t_{3}\pm i\omega
_{1}t_{1}},\label{eq:S_alpha}%
\end{equation}
obtained from the individual correlation functions $R_{\alpha}$ with
$t_{2}=0$ (i.e., at zero delay time). Spectra at nonzero delay could
be obtained by using $t_{2}>0$ in Eq.~(\ref{eq:S_alpha}). Spectra
$S_{\alpha}$ represent the ideal signals obtained in the limit of ultrashort pulses. In a more general setting with finite pulses, the
two-dimensional spectra are computed from the time-dependent polarization
(\ref{eq:P_t}), which involves explicitly the electric
fields.\cite{Schlau-Cohen_Fleming:2011,Do_Tan:2017} To ensure that all spectra
appear at positive frequencies $\omega_{1}$, nonrephasing spectra
($\alpha=1,4$) are computed with the positive sign in the exponent of
Eq.~(\ref{eq:S_alpha}), while the negative sign is used for the rephasing
spectra ($\alpha=2,3$).\cite{Schlau-Cohen_Fleming:2011} Furthermore, it is
easy to see from Eqs.~(\ref{eq:R1})--(\ref{eq:R4}) that $R_{1}(t_{3}%
,0,t_{1})=R_{4}(t_{3},0,t_{1})$ and $R_{2}(t_{3},0,t_{1})=R_{3}(t_{3}%
,0,t_{1})$; hence, we will show only two sets of spectra: $S_{1}\equiv S_{4}$
and $S_{2}\equiv S_{3}$. In general (i.e., for arbitrary $t_{2}%
$), correlation functions $R_{1}$ and $R_{2}$ are associated with the
stimulated emission process because the system evolves in the excited state
during the population time $t_{2}$; functions $R_{3}$ and $R_{4}$ correspond
to the ground-state bleaching because the system is in the ground electronic
state during the $t_{2}$ delay time. For $t_{2}=0$, one cannot
distinguish between these two processes.

To analyze the accuracy of different approximate approaches, we introduce the
spectral contrast angle
\begin{equation}
\cos\theta=\frac{S^{(\text{ref})}\cdot S}{\lVert S^{(\text{ref}) }\rVert\lVert
S \rVert}, \label{eq:cos_theta}%
\end{equation}
between the reference [$S^{(\text{ref})}$] and approximate ($S$) spectra, where
\begin{equation}
S^{(1)} \cdot S^{(2)}=\int d\omega_{1} \int d\omega_{3} S^{(1)}(\omega_{3},
\omega_{1}) S^{(2)}(\omega_{3}, \omega_{1})
\end{equation}
is the inner product of two two-dimensional spectra and $\lVert S \rVert
=\sqrt{S \cdot S}$ is the associated norm.

\section{\label{sec:compdet}Computational details}

\subsection{\label{subsec:morse_comp}One-dimensional models: Harmonic and
Morse potentials}

An arbitrary one-dimensional harmonic potential,
\begin{equation}
V_{\text{Harmonic}}(q; V_{\text{eq}}, q_{\text{eq}}, \omega) = V_{\text{eq}} +
\frac{1}{2} m \omega^{2} (q - q_{\text{eq}})^{2}, \label{eq:Harmonic_pot}%
\end{equation}
is described by the equilibrium position $q_{\text{eq}}$, energy minimum
$V_{\text{eq}}$, and frequency $\omega$. We set the mass $m = 1$ in all of our
model calculations. Let us also define a one-dimensional Morse potential,
\begin{equation}
V_{\text{Morse}}(q; V_{\text{eq}}, q_{\text{eq}}, \omega, \chi) =
V_{\text{eq}} + \frac{\omega}{4\chi}[1-e^{-\sqrt{2m\omega\chi}(q-q_{\text{eq}%
})}]^{2}, \label{eq:Morse_pot}%
\end{equation}
in terms of the anharmonicity parameter $\chi$ and the parameters
$V_{\text{eq}}$, $q_{\text{eq}}$, and $\omega$, which relate to the harmonic
potential (\ref{eq:Harmonic_pot}) fit to the Morse potential at $q_{\text{eq}%
}$.

We construct a set of one-dimensional systems composed of the ground-state
harmonic potential,
\begin{equation}
V_{1}(q) = V_{\text{Harmonic}}(q; V_{1, \text{eq}}=0, q_{1}=0, \omega_{1}=1)
\label{eq:ground_harmonic}%
\end{equation}
and the excited-state Morse potentials,
\begin{equation}
V_{2}(q)= V_{\text{Morse}}(q; V_{2, \text{eq}}=0, q_{2}=1.5, \omega_{2}=0.9,
\chi), \label{eq:excited_morse}%
\end{equation}
of variable anharmonicity $\chi$ ranging from $0.006$ to $0.02$. The initial
vibrational state, i.e., the ground vibrational state of the ground electronic
state, is a Gaussian due to the ground-state harmonic potential. The exact
two-dimensional electronic spectra are compared to approximate spectra
evaluated either with the harmonic approximation or with the thawed Gaussian
approximation. Within the harmonic approximation, the excited-state Morse
potential is replaced by the harmonic potential
\begin{equation}
V_{2}(q) \approx V_{\text{Harmonic}}(q; V_{2, \text{eq}}, q_{2}, \omega_{2}).
\label{eq:excited_harmonic}%
\end{equation}
Note that the harmonic result does not depend on the anharmonicity parameter
$\chi$ of the Morse potential.

Next, we compare the harmonic and thawed Gaussian approximations for a
one-dimensional system composed of two Morse potentials
\begin{align}
V_{1}(q)  &  =V_{\text{Morse}}(q;V_{1,\text{eq}},q_{1},\omega_{1}%
,\chi=0.01),\label{eq:ground_morse_0.01}\\
V_{2}(q)  &  =V_{\text{Morse}}(q;V_{2,\text{eq}},q_{2},\omega_{2},\chi=0.01),
\label{eq:excited_morse_0.01}%
\end{align}
with the same degree of anharmonicity. The exact initial state is no more a
Gaussian. However, in the thawed Gaussian simulations, we approximate it by
the vibrational ground state of the harmonic potential
(\ref{eq:ground_harmonic}) fitted to the ground-state Morse potential
(\ref{eq:ground_morse_0.01}) at its minimum. The harmonic approximation
replaces both ground-state and excited-state potential energy surfaces by the 
harmonic potentials; the result is the same as for the harmonic-Morse system
described above.

Wavepacket propagation was performed for 150 steps in both $t_{1}$ and $t_{3}$
times and with a time step of 0.2. The transition dipole moment was set to 1
(Condon approximation). Correlation functions $R_{1}$, $R_{2}$, $R_{3}$, and $R_{4}$ were multiplied
by a Gaussian damping function $\exp[-a(t_{1}^{2} + t_{3}^{2})]$ with $a =
0.014427$, resulting in the Gaussian broadening (half-width at half-maximum of
0.2) of the spectra along both frequency axes. The exact spectra were computed
in the eigenstate representation, which is feasible for these one-dimensional
systems since both harmonic and Morse eigenfunctions are
known;\cite{Morse:1929} the associated Franck-Condon overlaps were computed numerically.

\subsection{\label{subsec:otf_comp}On-the-fly \textit{ab initio} calculations}

The electronic structure of phenol was modeled using the density functional
theory with the PBE0 functional and 6-311G(d, p) basis set, as implemented in
the Gaussian 16 quantum chemistry package.\cite{Frisch_Fox:2016} Excited-state
calculations were performed with the time-dependent density functional theory.
This choice of electronic structure theory provides ground-state frequencies
similar to those computed at the MP2/aug-cc-pVDZ level (see Table~II of the
supplementary material and Ref.~\onlinecite{Rajak_Mahapatra:2018}) and
transition energies along the excited-state trajectory that agree, up to
an approximately constant shift (which results only in a shift of the
computed spectrum but does not affect its shape), to those evaluated at the EOM-CCSD/6-311G(d, p) level (Fig. 1 of the
supplementary material). A single ab initio excited-state trajectory was run
for 1000 steps starting from the ground-state optimized geometry; subsequent
ground-state classical trajectories were propagated for 500 steps. Overall,
the calculations allow the evaluation of the correlation functions with 500
steps in both $t_{1}$ and $t_{3}$ delay times; $t_{2}$ delay was set to zero.
All dynamics simulations used a time step of $0.25\,$fs and a standard
second-order Verlet integrator. The ab initio calculations evaluated not only
the energies and gradients at each step but also the Hessians of the
electronic energy. These potential energy data were transformed to
ground-state normal mode coordinates and used to propagate the 33-dimensional wavepacket
according to Eqs.~(\ref{eq:q_t_dot})--(\ref{eq:gamma_t_dot}). After evolving
the ground- and excited-state Gaussian wavepackets, the correlation functions
were computed using Eqs.~(\ref{eq:R1})--(\ref{eq:R4}), (\ref{eq:C_wp}), and the expression
\begin{equation}
\langle\psi_{1}|\psi_{2}\rangle=\sqrt{\frac{(2\pi\hbar)^{D}}{\det(-i\delta
A)}}\exp\left\{  \frac{i}{\hbar}\left[  -\frac{1}{2}\delta\xi^{T}\cdot(\delta
A)^{-1}\cdot\delta\xi+\delta\eta\right]  \right\}  \label{eq:psi_overlap}%
\end{equation}
for the overlap of two thawed Gaussian wavepackets with parameters
$q_{i}$, $p_{i}$, $A_{i}$, and $\gamma_{i}$ ($i=1,2$). In Eq.~(\ref{eq:psi_overlap}), we defined
vectors and scalars
\begin{align}
\xi_{i} &  :=p_{i}-A_{i}\cdot q_{i},\\
\eta_{i} &  :=\gamma_{i}-\frac{1}{2}(\xi_{i}+p_{i})^{T}\cdot q_{i},
\end{align}
as well as the notation $\delta\Lambda:=\Lambda_{2}-\Lambda_{1}%
^{\ast}$ for $\Lambda=A$, $\xi$, $\eta$.

To construct the harmonic model, also known as the generalized Brownian
oscillator model,\cite{Zuehlsdorff_Isborn:2019} of phenol, an additional
Hessian was computed at the optimized excited-state geometry. This corresponds
to the so-called adiabatic Hessian or adiabatic harmonic
model.\cite{AvilaFerrer_Santoro:2012,Vanicek_Begusic:2021} Two more
approximate models were also studied: The uncoupled harmonic model was
obtained by neglecting the off-diagonal terms of the excited-state Hessian
expressed in terms of the ground-state normal modes. The displaced harmonic
oscillator model, also called the Brownian oscillator model, was constructed
by replacing the excited-state Hessian in the adiabatic harmonic model by the
ground-state Hessian; this specific way of constructing the displaced harmonic
oscillator parameters is called the adiabatic shift
approach.\cite{AvilaFerrer_Santoro:2012,Vanicek_Begusic:2021} When applied to any of these different harmonic potentials, the thawed Gaussian propagation is exact and enables an efficient evaluation of linear and two-dimensional spectra. Although explicit expressions are available for the evaluation of linear absorption and emission spectra of harmonic systems,\cite{Baiardi_Barone:2013} no such analytical approaches have been presented for the two-dimensional spectra of arbitrarily shifted, distorted, and rotated harmonic potentials.

Spectra simulations assumed Condon approximation for the transition dipole
moment. Linear absorption spectra were computed from the first 500 steps of the excited-state wavepacket autocorrelation function (see Fig.~2 of the supplementary material, where the convergence is confirmed, and Ref.~\onlinecite{Vanicek_Begusic:2021} for more details) and were broadened by a Gaussian with half-width
at half-maximum of $120\,$cm$^{-1}$; same broadening was used for the
two-dimensional spectra along both $\omega_{1}$ and $\omega_{3}$ frequency
axes. This corresponds to a phenomenological inhomogeneous broadening;
homogeneous broadening due to direct system-bath interactions is neglected.
The system-bath coupling would be needed for spectra at later delay times
$t_{2} > 0$, as the system would have time to relax and dissipate energy to
the environment; we assume that the response functions with $t_{2} = 0$ and
$t_{1}, t_{3} < 125\,$fs are only weakly affected by the system-bath coupling.

\section{\label{sec:resanddisc}Results and discussion}

\subsection{\label{subsec:model_res}Model potentials}

\subsubsection{\label{subsubsec:harmonic_morse} Harmonic-Morse system}

\begin{figure}
\includegraphics[width = \textwidth]{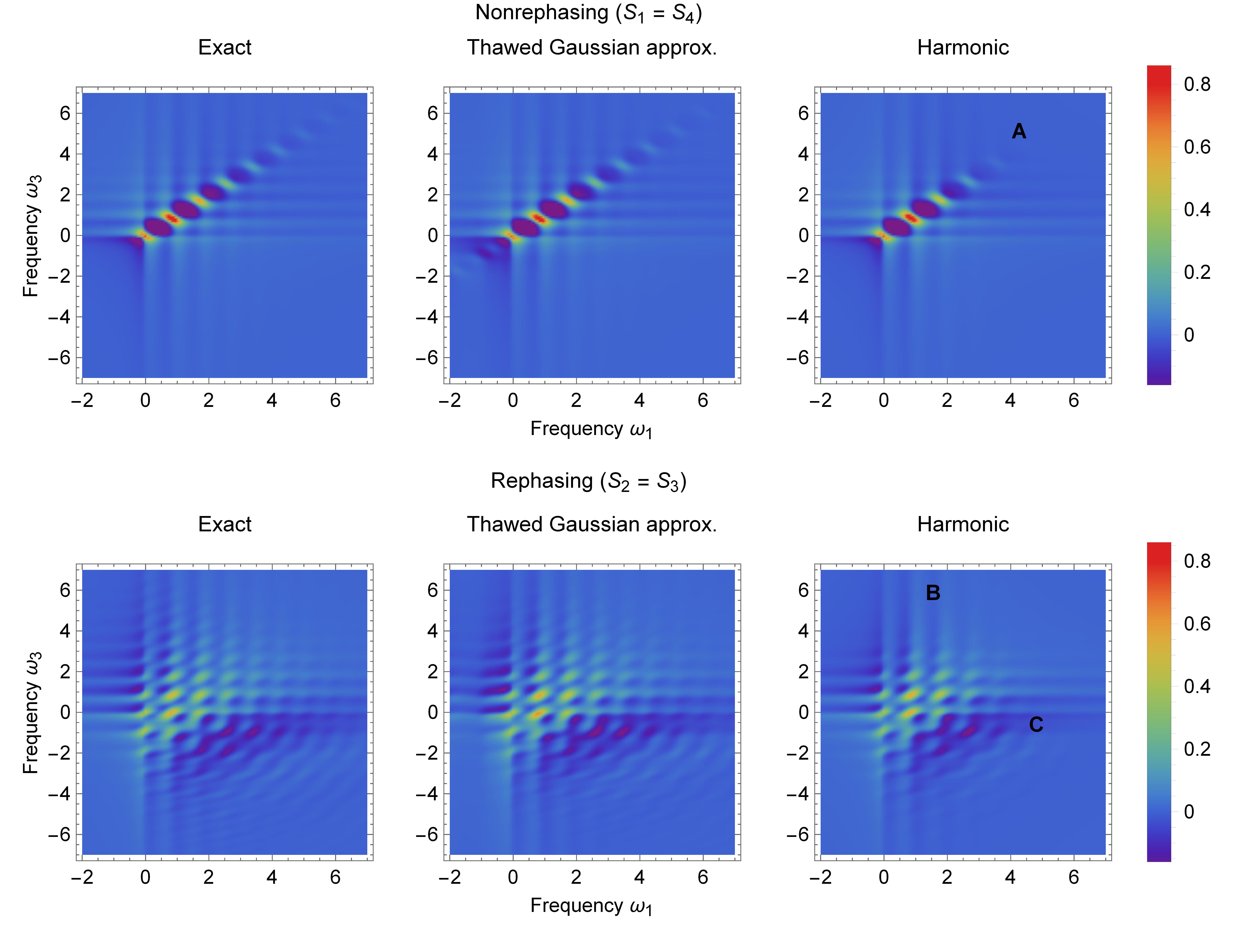} \caption{\label{fig:Spectra2DMorse}Exact, thawed Gaussian, and harmonic two-dimensional electronic spectra for the harmonic-Morse system described in Sec.~\ref{subsec:morse_comp} with the anharmonicity of the excited-state Morse potential $\chi = 0.01$. Spectral regions A, B, and C, discussed in the main text, are indicated on the harmonic spectra.}
\end{figure}

Two-dimensional spectra for the harmonic ground-state potential and Morse
excited-state potential are shown in Fig.~\ref{fig:Spectra2DMorse}. The exact
nonrephasing spectrum appears only along the diagonal, whereas the rephasing
spectrum exhibits a characteristic checkerboard pattern due to vibronic
transitions that involve various ground- and excited-state vibrational states.
Already at first sight, it is clear that the spectra evaluated within the
thawed Gaussian approximation reproduce the exact spectra well, which is not
the case for the harmonic results. The nonrephasing harmonic spectrum flattens
out at higher frequencies (spectral region A indicated in the top right panel
of Fig.~\ref{fig:Spectra2DMorse}), unlike the exact and thawed Gaussian
spectra, which exhibit clear vibronic peaks at these frequencies. Similar
effects are seen in the rephasing spectra, mostly in the spectral region
labeled B (see Fig.~\ref{fig:Spectra2DMorse}, bottom right). Again, the exact
spectrum is composed of a long vibronic progression up to $\omega_{3}=6$,
which, in the harmonic approximation, is truncated around $\omega_{3}=4$. In
the region C, the harmonic spectrum is missing negative vibronic peaks, which
are reproduced well by the thawed Gaussian approximation. The thawed
Gaussian approximation, however, suffers from another form of error: as in
linear spectroscopy (see, e.g., Ref.~\onlinecite{Wehrle_Vanicek:2015}),
artificial negative peaks may also appear in the two-dimensional
spectra, which is most obvious in the nonrephasing spectrum of
Fig.~\ref{fig:Spectra2DMorse} around $(\omega_{1},\omega_{3})\approx(-1,-1)$.

\begin{figure}
\includegraphics[width = 0.5\textwidth]{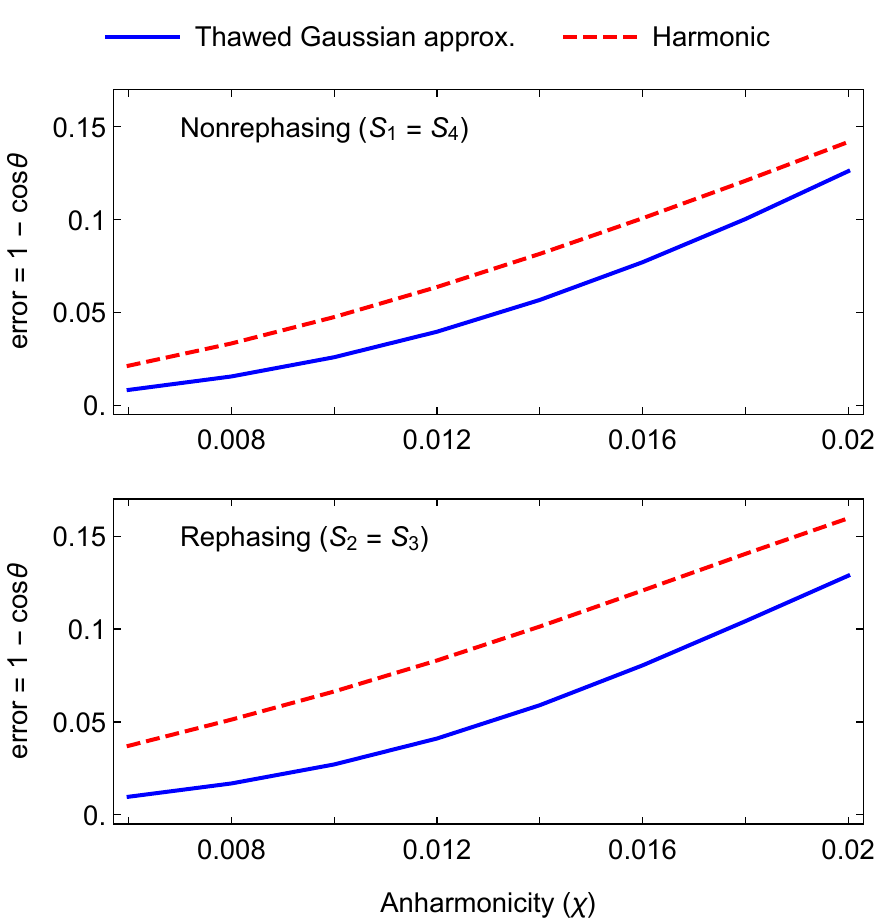} \caption{\label{fig:Errors2DMorse} Errors of the thawed Gaussian and harmonic spectra of the harmonic-Morse system, measured by the spectral contrast angles [Eq.~(\ref{eq:cos_theta})] at different values of the anharmonicity parameter $\chi$.}
\end{figure}

We now compare the exact and approximate spectra at different levels of
anharmonicity by measuring the error (see Fig.~\ref{fig:Errors2DMorse})
through the spectral contrast angle [Eq.~(\ref{eq:cos_theta})]. The thawed
Gaussian approximation exhibits smaller errors in the computed spectra than
the harmonic approximation at all levels of anharmonicity and for both
rephasing and nonrephasing spectra. As expected, the accuracy of both
approximate approaches deteriorates as the anharmonicity of the system increases.

\subsubsection{\label{subsubsec:morse_morse} Morse-Morse system}

\begin{figure}
\includegraphics[width = \textwidth]{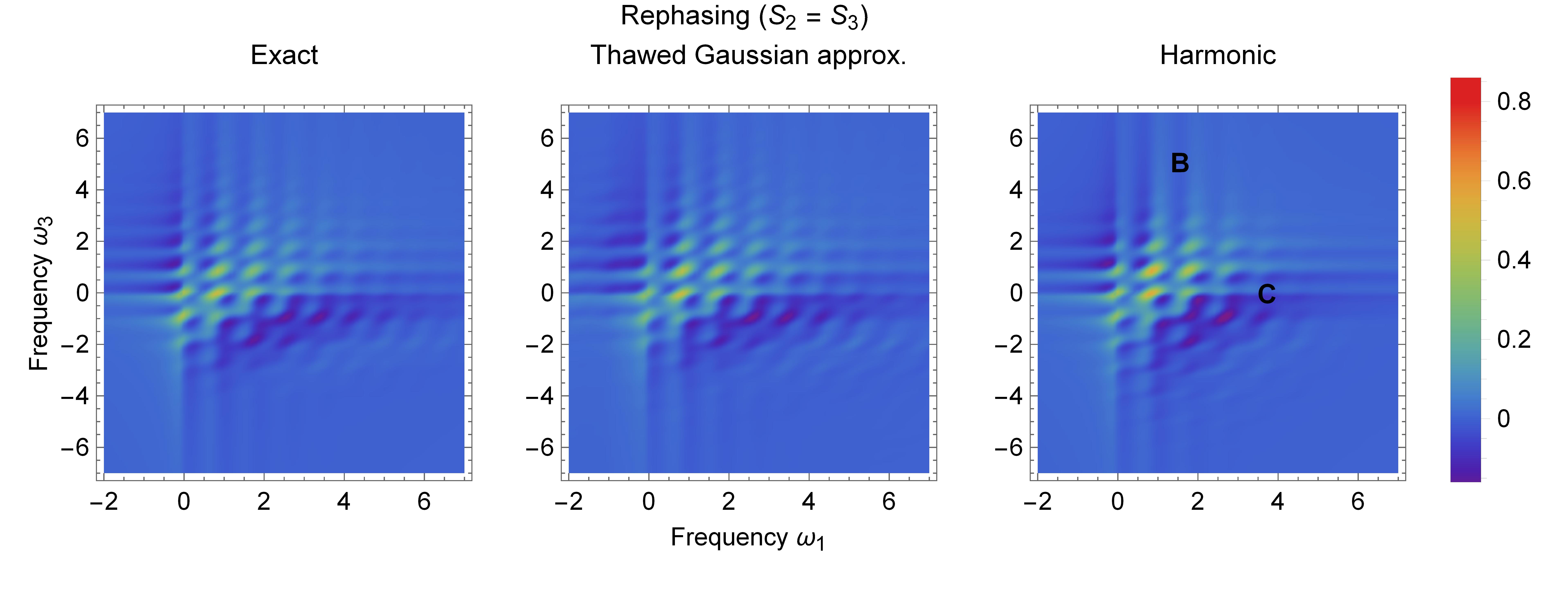} \caption{\label{fig:Spectra2DMorse_Morse}Exact, thawed Gaussian, and harmonic rephasing spectra of the Morse-Morse system described in Sec.~\ref{subsec:morse_comp} (anharmonicity $\chi = 0.01$).}
\end{figure}

\begin{figure}
\includegraphics[width = 0.5\textwidth]{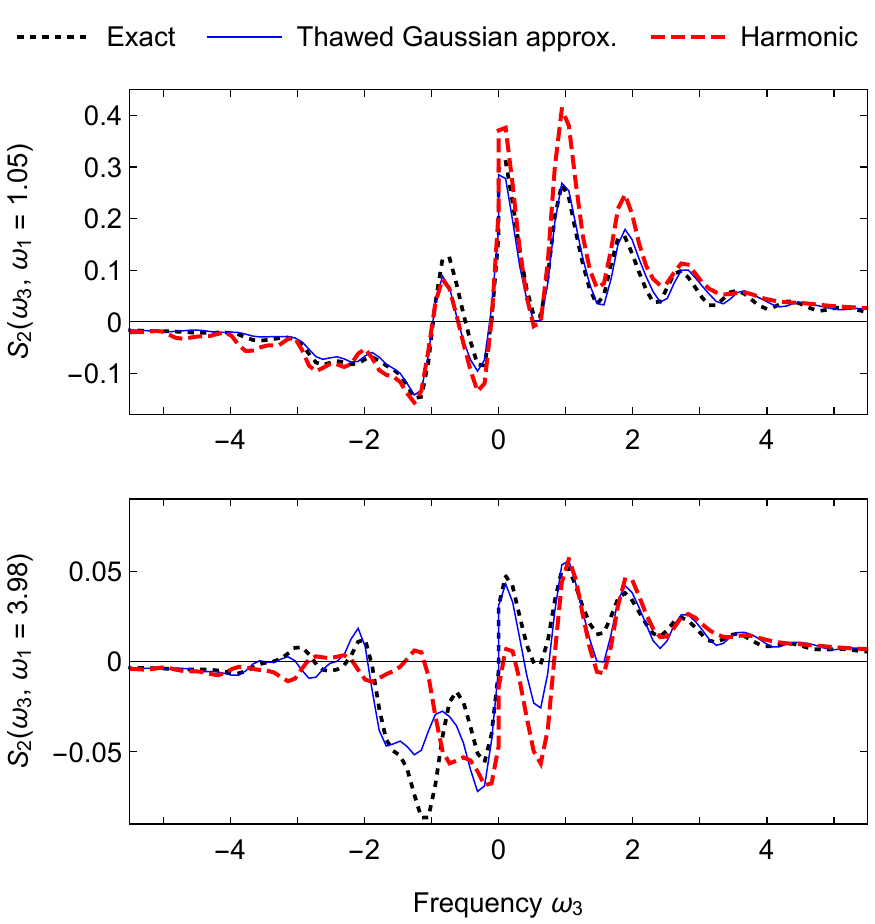} \caption{\label{fig:CutsMorse_Morse}One-dimensional cuts of the two-dimensional spectra of Fig.~\ref{fig:Spectra2DMorse_Morse} at $\omega_1 \approx 1$ (top) and $\omega_1 \approx 4$ (bottom).}
\end{figure}

Two-dimensional rephasing spectra of the system composed of two Morse
potentials, both with the anharmonicity parameter $\chi= 0.01$, are shown in
Fig.~\ref{fig:Spectra2DMorse_Morse}. As in the harmonic-Morse system, the
errors of the harmonic spectrum are observed in the spectral regions B and C;
the accuracy of the thawed Gaussian spectrum is not much affected by the
additional anharmonicity in the ground-state potential surface. To analyze
further the two approximate methods, we inspect one-dimensional cuts of the
two-dimensional spectra along two different values of $\omega_{1}$ frequency
(Fig.~\ref{fig:CutsMorse_Morse}). We see clearly that the thawed Gaussian
approximation recovers the positions and intensities of the vibronic peaks
both at low $\omega_{1} \approx1$ and high $\omega_{1} \approx4$ frequencies.
Harmonic results recover qualitatively the spectral cut at the lower
$\omega_{1}$ frequency (Fig.~\ref{fig:CutsMorse_Morse}, top) but fail to
recover the vibronic peaks at the higher $\omega_{1}$ frequency
(Fig.~\ref{fig:CutsMorse_Morse}, bottom). Notably, the negative peak at
$\omega_{3} \approx-1$ is missing in the spectrum calculated within the
harmonic approximation. Such errors could, in practice, seriously affect the
interpretation of the experiments. One of the main challenges in
two-dimensional electronic spectroscopy is to assign spectral features to
either vibrational or electronic degrees of
freedom.\cite{Butkus_Abramavicius:2012,Butkus_Abramavicius:2012a,Turner_Scholes:2011}
If the simulation, for example, based on a model harmonic potential, cannot
reproduce the vibronic peaks found in the experimental spectra, these peaks
might end up incorrectly assigned to another electronic state or another
excitation process.

\subsection{\label{subsec:phenol}Two-dimensional electronic spectrum of phenol}

\begin{figure}
\includegraphics[width = 0.5\textwidth]{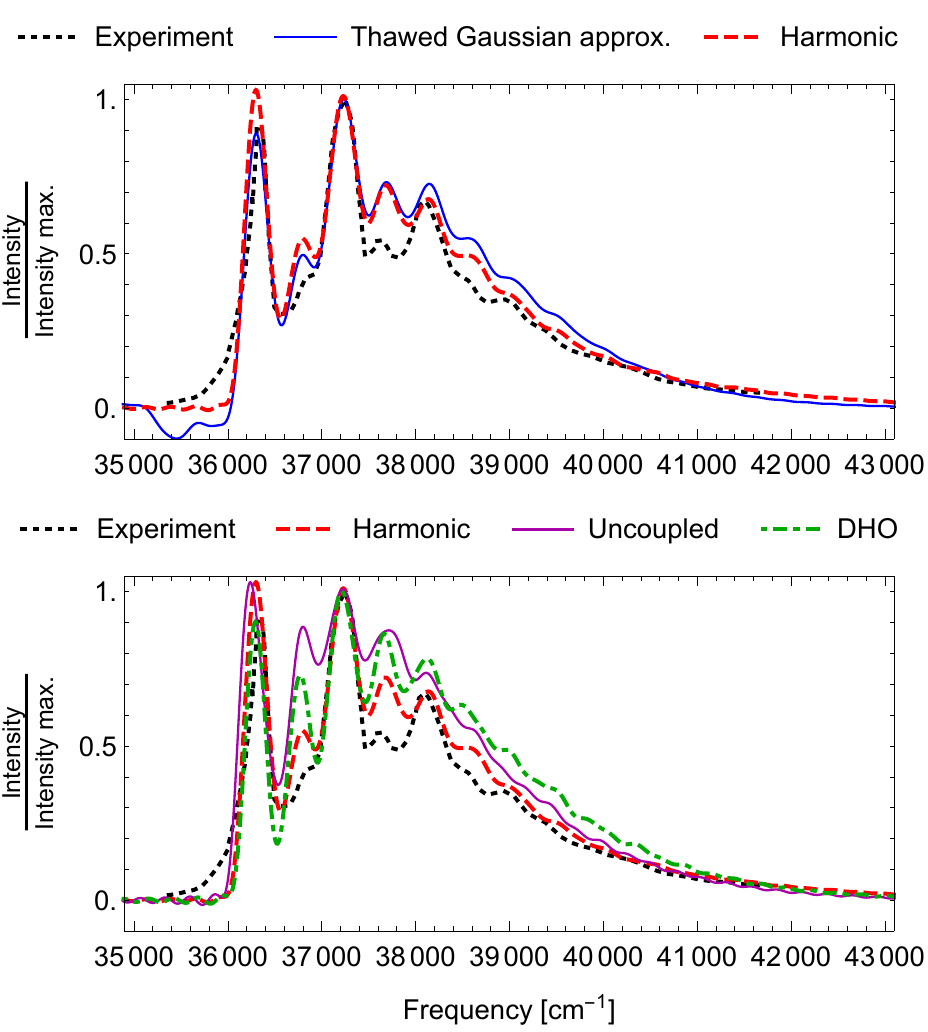}
\caption{\label{fig:LinearSpectraFC}Experimental linear absorption spectrum of
Ref.~\onlinecite{Karmakar_Chakraborty:2015} (data extracted with WebPlotDigitizer\cite{webplotdigitizer}) and the spectra computed with the on-the-fly ab initio thawed
Gaussian approximation, harmonic approximation, uncoupled harmonic model (``Uncoupled''), and displaced
harmonic oscillator (DHO) model. For ease of comparison, the computed spectra are shifted in frequency and
scaled in intensity so that they all match at the maximum of the experimental spectrum (see Sec.~IV of the
supplementary material for details).}
\end{figure}

Phenol is an ultraviolet chromophore present in proteins as the residue of the
naturally occurring amino acid tyrosine. Recently, accurate electronic
structure methods were employed to simulate its two-dimensional electronic
spectrum\cite{Nenov_Rivalta:2015,Segarra-Marti_Rivalta:2018} in an attempt to
explore theoretically the capabilities of this spectroscopic technique to
resolve the features specific to chromophore-chromophore interactions in
oligopeptides and, more generally, in
proteins.\cite{Nenov_Garavelli:2014,Giussani_Nenov:2017} These recent
calculations included multiple electronic states but neglected the vibronic
structure of the individual electronic transitions. Here, we present a
complementary result: we focus only on the ground and first excited electronic
states, i.e., we neglect the excited-state absorption process, but study in
detail the vibronic lineshape of the ground-state bleaching/stimulated
emission signal. The methods we use neglect the nonadiabatic effects; this is
an acceptable approximation for the dynamics in the first excited state of
phenol, as demonstrated by the MCTDH simulations performed on a
vibronic-coupling Hamiltonian model of phenol.\cite{Rajak_Mahapatra:2018}

The linear absorption spectrum of phenol was computed with four different
approximate methods: the on-the-fly ab initio thawed Gaussian approximation,
harmonic approximation, uncoupled harmonic model, and displaced harmonic
oscillator model (Fig.~\ref{fig:LinearSpectraFC}). Harmonic and on-the-fly
thawed Gaussian spectra (Fig.~\ref{fig:LinearSpectraFC}, top) are similar in
accuracy for this specific system: while the thawed Gaussian propagation
results in more accurate intensities of the low-frequency peaks, namely, the 0--0 transition at $\approx 36350\ $cm$^{-1}$ and the shoulder at $\approx 36800\ $cm$^{-1}$, the harmonic approximation gives
a better estimate of the high-frequency region and the tail of the spectrum. One of the main disadvantages
of the thawed Gaussian approximation, the appearance of artificial negative
spectral intensities, shows up clearly in the absorption spectrum of phenol. Although the simulated harmonic and thawed Gaussian spectra resemble the experiment, there are remaining differences, most notably in the intensities of the spectral peaks. These errors could be either due to anharmonicity effects not captured by the approximate thawed Gaussian wavepacket propagation or due to the errors in the potential energy data evaluated with an approximate electronic structure method. Fairly small difference between the harmonic and thawed Gaussian spectra suggests that the anharmonicity effects are weak and that the remaining errors in our simulation are due to the inaccuracies of the electronic structure theory. In the bottom panel of Fig.~\ref{fig:LinearSpectraFC}, we show the results of two more approximate approaches---the uncoupled harmonic and displaced
harmonic oscillator models. These spectra clearly deviate from the
experiment, indicating the importance
of both mode distortion (changes in mode frequencies) and intermode couplings
(Duschinsky effect). When going from the displaced harmonic model, which
neglects mode distortion, to the uncoupled harmonic model, which includes mode
distortion, the peaks broaden but still exhibit inaccurate intensities.
Additional inclusion of the Duschinsky effect, which is achieved by moving to
the (coupled) harmonic model, improves the intensities.

\begin{figure}
\includegraphics[width = 0.55\textwidth]{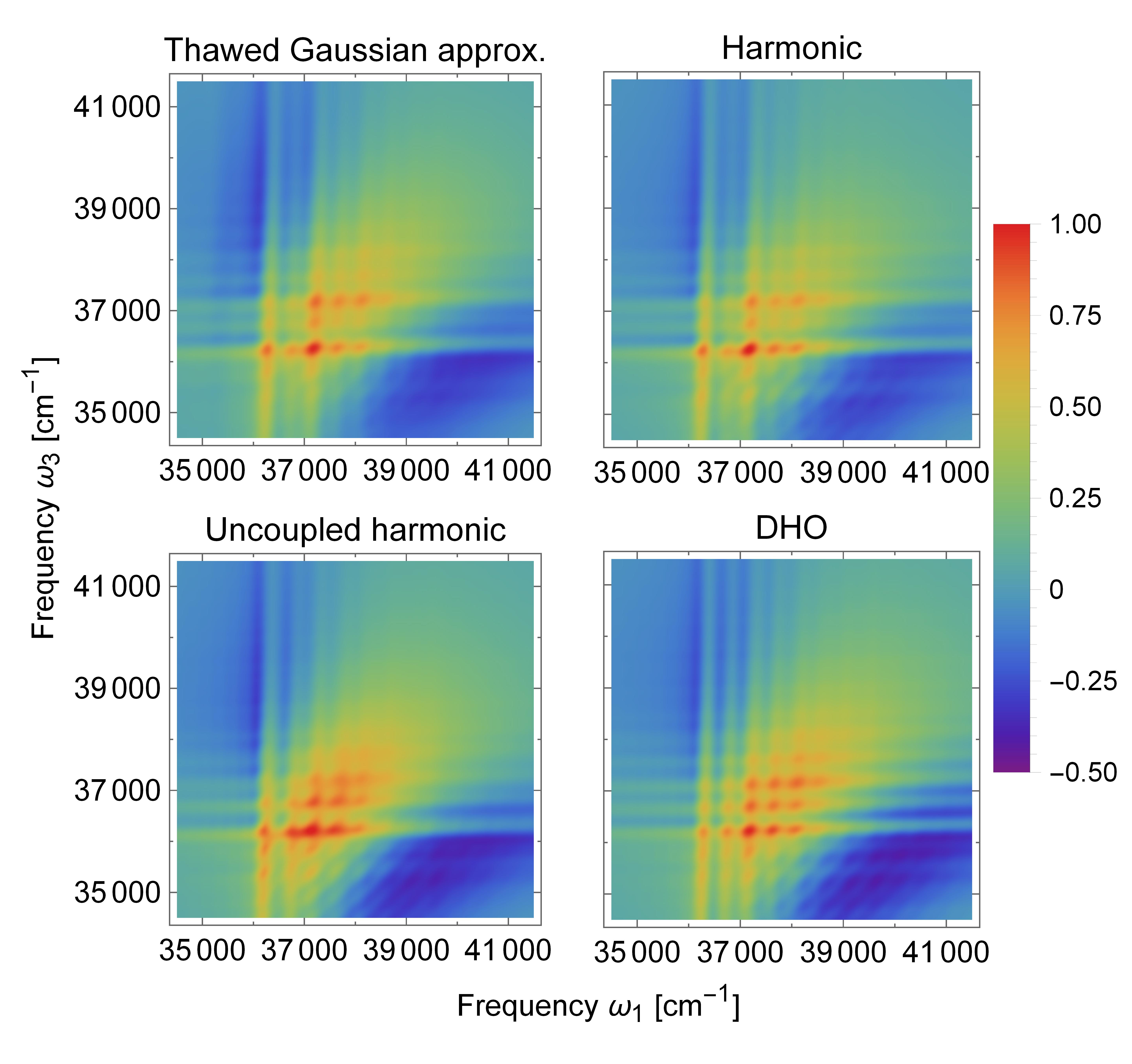}
\caption{\label{fig:Phenol2D}Rephasing two-dimensional electronic spectra of phenol at $t_2 = 0$ computed with
the on-the-fly ab initio thawed Gaussian approximation, harmonic approximation, uncoupled harmonic model, and
displaced harmonic oscillator (DHO) model. The spectra correspond to the stimulated emission ($S_2$) and
ground-state bleaching ($S_3$) processes ($S_2 = S_3$ for $t_2 =0$). Computed spectra were shifted along
both frequency axes and scaled in intensity as in Fig.~\ref{fig:LinearSpectraFC}.}
\end{figure}

The two-dimensional spectra simulated with different approximate methods are
shown in Fig.~\ref{fig:Phenol2D}. Again, the uncoupled harmonic and displaced
harmonic oscillator models predict spectra that differ substantially from the
harmonic and thawed Gaussian results, which are, in turn, similar to each
other. Therefore, based on both linear and two-dimensional spectra
simulations, we may conclude that the anharmonicity effects are truly weak in
the ground and first excited states of phenol, at least in the region explored
by the nuclear wavepacket for short time after the photoexcitation. More precisely, the anharmonicity effects are much weaker than the effects of the
Duschinsky rotation and frequency changes, which are, in contrast, significant, as
demonstrated by the simulations based on the uncoupled or displaced harmonic
models. The anharmonicity could, however, play a role at longer simulation times, needed, for example, to simulate high-resolution spectra. We note that most simulations supporting experimental results are
nowadays performed with the simplified displaced harmonic oscillator model,
which captures the basic physics of the problem, but is inadequate in certain
cases, such as the presented example of phenol.

Interestingly, the spectrum spans a broad range of frequencies in both
$\omega_{1}$ and $\omega_{3}$, which is in stark contrast with the simulations
of Ref.~\onlinecite{Nenov_Rivalta:2015}. More specifically, the broad vibronic
ground-state bleach/stimulated emission spectrum is expected to overlap
strongly with the excited-state absorption signals of phenol and even with the
signals of other amino acid residues (compare our results with those for a
noninteracting benzene-phenol dimer in Fig.~3 of
Ref.~\onlinecite{Nenov_Rivalta:2015}). Hence, an accurate treatment of
vibronic effects is needed to simulate realistic spectra and to help explain
these overlapping, unresolved spectral features. Our results also support
indirectly the concluding part of Ref.~\onlinecite{Nenov_Rivalta:2015}, where
a two-color ultraviolet-visible experiment is proposed to resolve transitions
to charge-transfer states (see Fig.~6 of
Ref.~\onlinecite{Nenov_Rivalta:2015}), which appear only when the two
chromophores are close to each other. In the visible region of frequency
$\omega_{3}$, there are fewer spectroscopic transitions and these charge
transfer states could be easily distinguished from the states of the
individual chromophores even with broad vibronic features included.

\section{Conclusions and outlook}

We have presented a new method for simulating vibrationally resolved
two-dimensional electronic spectra that is exact for any shifted,
distorted, and coupled harmonic model and, in addition, can approximately
account for anharmonicity effects. The method, based on the thawed
Gaussian approximation, is shown to be superior to the harmonic approximation
for a series of Morse models of varying anharmonicity. On the example of
phenol, we show that inter-mode couplings and changes in the mode frequencies,
both of which are frequently neglected in simulations, can be crucial
for recovering the correct vibronic shape of the two-dimensional electronic
spectra. In this specific case, the anharmonicity is shown to be weak, which
could allow further studies on the nonlinear spectra of phenol based on the
harmonic approximation. For example, our results could be augmented by
constructing harmonic models with more accurate electronic structure methods,
in order to simulate excited-state absorption signals. For systems that do
exhibit anharmonicity effects, we propose the on-the-fly ab initio thawed
Gaussian approximation as a computationally affordable approach beyond
harmonic approximation.

Finally, let us also give a short outlook on how to include features that are
missing in the current method. First, as a wavefunction method, the thawed
Gaussian approximation is not suitable for treating systems at non-zero
temperature. We have shown recently that this limitation can be overcome
efficiently with the so-called thermo-field dynamics
theory.\cite{Begusic_Vanicek:2020} Currently, we are exploring the application
of this idea to the computation of nonlinear spectra. Second, the method is
originally constructed for isolated systems. An obvious, \textquotedblleft ab
initio way\textquotedblright\ to augment the system with an environment would
be to include a number of solvent molecules directly into the system. To
account for inhomogeneous broadening, the dynamics would have to be repeated
for different conformations of the solute-solvent system. Alternatively, the
bath effects could be treated through a number of low-frequency
harmonic oscillators coupled to the system; the procedures for computing the
parameters of the bath oscillators are well-studied in the literature. The
extensions that include temperature and environment effects would enable
accurate and efficient first-principles simulation of time-resolved ($t_{2}%
>0$) two-dimensional electronic spectra in the condensed phase.

\section*{Supplementary material\label{sec:supp_mat}}

See the supplementary material for ground- and excited-state
optimized geometries, normal-mode frequencies and displacements, validation of
the electronic structure method, wavepacket autocorrelation function, and frequency shifts applied to the computed
spectra of phenol. Supplementary material contains Refs.~\onlinecite{Gao_Freindorf:1996,Granucci_Tran-Thi:2000,Rogers_Hirst:2003,Lan_Mahapatra:2005,Zhang_Muchall:2006,Vieuxmaire_Domcke:2008,Kim_Kang:2009,Yang_Truhlar:2014}.

\begin{acknowledgments}
The authors acknowledge the financial support from the Swiss National Science
Foundation through the NCCR MUST (Molecular Ultrafast Science and Technology)
Network and from the European Research Council (ERC) under the European
Union's Horizon 2020 research and innovation programme (grant agreement No.
683069 -- MOLEQULE).
\end{acknowledgments}

\section*{Data availability\label{sec:data}}

Data that support the findings of this study are openly available in Zenodo at\\ http://doi.org/10.5281/zenodo.4121622.

\bibliographystyle{aipnum4-2}
\bibliography{biblio48,additions_ThirdOrder}

\end{document}

\title{Supplementary material for: On-the-fly \textit{ab initio} semiclassical evaluation of third-order response functions for two-dimensional electronic spectroscopy}
\author{Tomislav Begu\v{s}i\'{c}}
\email{tomislav.begusic@epfl.ch}
\author{Ji\v{r}\'i Van\'i\v{c}ek}
\email{jiri.vanicek@epfl.ch}
\affiliation{Laboratory of Theoretical Physical Chemistry, Institut des Sciences et
Ing\'enierie Chimiques, Ecole Polytechnique F\'ed\'erale de Lausanne (EPFL),
CH-1015, Lausanne, Switzerland}
\date{\today}

\begin{abstract}

\end{abstract}

\maketitle

\graphicspath{{"C:/Users/GROUP LCPT/Documents/Group/Tomislav/ThirdOrder/figures/"}
{./figures/}{C:/Users/Jiri/Dropbox/Papers/Chemistry_papers/2020/ThirdOrder/figures/}}

\section{Ground- and excited-state optimized geometries, frequencies, and relative displacements}

\begin{table}
[H]%
\caption{Optimized geometries (in \AA) of the ground and first excited electronic states of phenol at the PBE0/6-311G(d, p) level of theory.}
\centering
\renewcommand{\arraystretch}{1} \begin{ruledtabular}
\begin{tabular}{ccccccccc}
\multicolumn{3}{r}{S$_0$} &&& \multicolumn{3}{r}{S$_1$} \\
&X 	  &Y 	   &Z 		&&&X 	  &Y 	   &Z\\ \hline
C & -0.000  &  0.940  &  0.000  &&& -0.930  &  0.038 & -0.008\\
C & -1.199  &  0.232  &  0.000  &&& -0.298  & -1.235 & -0.019\\
C & -1.183  & -1.158  &  0.000  &&&  1.120  & -1.261 &  0.037\\
C &  0.021  & -1.849  &  0.000  &&&  1.831  & -0.039 &  0.010\\
C &  1.215  & -1.134  &  0.000  &&&  1.190  &  1.218 & -0.036\\
C &  1.212  &  0.252  &  0.000  &&& -0.230  &  1.264 &  0.006\\
O &  0.049  &  2.296  &  0.000  &&& -2.269  &  0.119 &  0.002\\
H & -0.848  &  2.639  &  0.000  &&& -2.630  & -0.770 &  0.097\\
H & -2.146  &  0.767  &  0.000  &&& -0.882  & -2.132 & -0.199\\
H & -2.123  & -1.700  &  0.000  &&&  1.656  & -2.201 &  0.062\\
H &  0.031  & -2.933  &  0.000  &&&  2.917  & -0.072 &  0.018\\
H &  2.163  & -1.662  &  0.000  &&&  1.773  &  2.129 & -0.061\\
H &  2.134  &  0.822  &  0.000  &&& -0.787  &  2.185 &  0.126
\end{tabular}
\end{ruledtabular}
\label{tab:Phenol_optgeom}
\end{table}

\begin{table}[H]
\caption{Ground- and excited-state frequencies and dimensionless relative displacements between the two states. Frequencies are given in $\text{cm}^{-1}$. Dimensionless displacements are defined as $|\sqrt{\omega_i/\hbar} q_{2,i}|$, where $\omega_i$ is the $i$-th ground-state frequency and $q_{2,i}$ is the excited-state position in the mass-scaled normal mode coordinate $i$ of the ground state. Both ground- and excited-state frequencies are listed in the descending order.}
\label{tab:Phenol_freq} \centering
\renewcommand{\arraystretch}{0.65} \begin{ruledtabular}
\begin{tabular}{ccccccccccccc}
&Mode & S$_0$ frequencies & S$_1$ frequencies & Displacement \\
\hline
& 1  & 3889 & 3806 & 0.012 \\
& 2  & 3222 & 3242 & 0.127 \\
& 3	 & 3216 & 3232 & 0.041 \\
& 4  & 3201 & 3225 & 0.079 \\
& 5  & 3192 & 3193 & 0.066 \\
& 6	 & 3171 & 3182 & 0.165 \\
& 7  & 1682 & 1579 & 0.215 \\
& 8  & 1668 & 1499 & 0.077 \\
& 9	 & 1544 & 1492 & 0.143 \\
& 10 & 1511 & 1468 & 0.027 \\
& 11 & 1388 & 1439 & 0.085 \\
& 12 & 1361 & 1340 & 0.082 \\
& 13 & 1314 & 1320 & 0.553 \\
& 14 & 1207 & 1181 & 0.016 \\
& 15 & 1190 & 1159 & 0.035 \\
& 16 & 1175 & 1148 & 0.028 \\
& 17 & 1100 & 1022 & 0.113 \\
& 18 & 1055 & 996  & 0.587 \\
& 19 & 1015 & 987  & 0.708 \\
& 20 & 989  & 854  & 0.081 \\
& 21 & 969  & 822  & 0.360 \\
& 22 & 890  & 678  & 0.044 \\
& 23 & 840  & 631  & 0.819 \\
& 24 & 825  & 608  & 0.724 \\
& 25 & 766  & 542  & 0.174 \\
& 26 & 702  & 527  & 0.079 \\
& 27 & 631  & 519  & 0.113 \\
& 28 & 536  & 498  & 0.972 \\
& 29 & 519  & 400  & 0.003 \\
& 30 & 421  & 345  & 0.893 \\
& 31 & 409  & 297  & 0.111 \\
& 32 & 352  & 187  & 0.357 \\
& 33 & 232  & 117  & 0.218
\end{tabular}
\end{ruledtabular}

\end{table}

\section{Equation-of-motion coupled cluster singles and doubles (EOM-CCSD) transition energies along the excited-state trajectory}

\begin{figure}[H]
\centering
\includegraphics[scale=1]{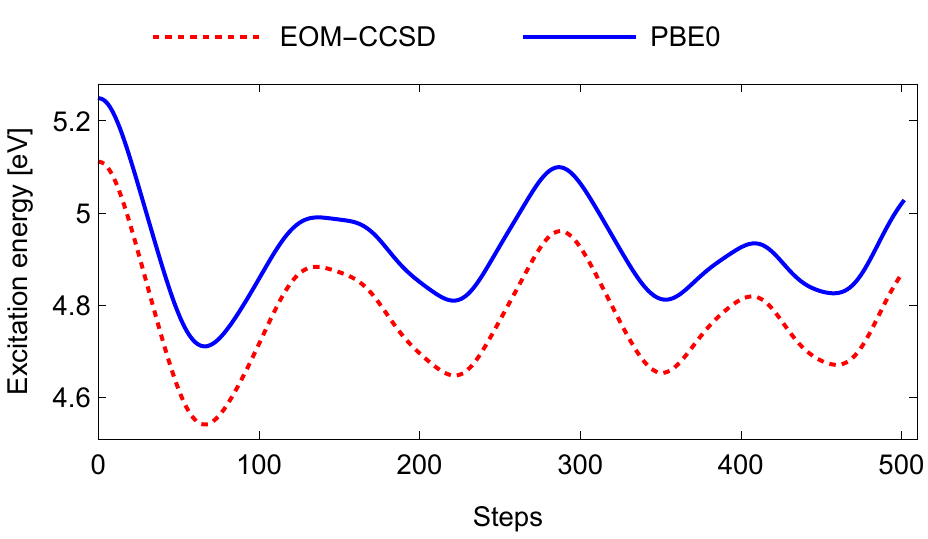}
\caption{The energy gaps between the excited and ground electronic states of phenol evaluated along the first 500 steps of the excited-state \textit{ab initio} trajectory at the PBE0/6-311G(d, p) level of theory, compared to the transition energies computed using the EOM-CCSD/6-311G(d, p) method. The nearly constant shift in the transition energies induces only a constant frequency shift in the computed spectra, but does not affect their shape.}
\label{fig:ExcitationEnergies}
\end{figure}

\section{Wavepacket autocorrelation function for the linear absorption spectrum}

\begin{figure}[H]
\centering
\includegraphics[scale=1]{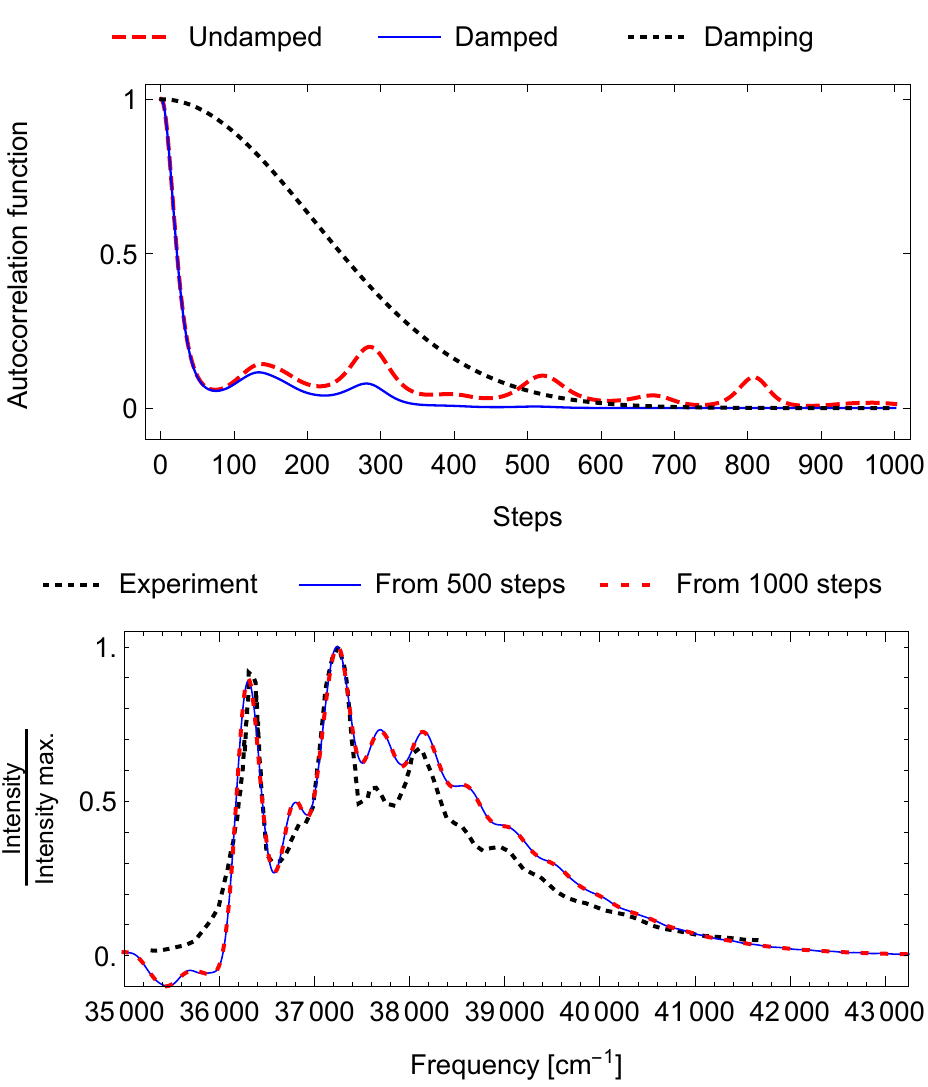}
\caption{Top: Undamped (red, dashed) and damped (blue, solid) wavepacket autocorrelation function $\langle 1, g | e^{- i \hat{H}^{\prime}_2 t / \hbar} | 1, g\rangle$, evaluated with the on-the-fly \textit{ab initio} thawed Gaussian approximation, and the damping (black, dotted), which is a Gaussian function $\exp(-t^2/2\sigma_t^2)$ with $\sigma_t = \sqrt{2 \ln 2} / \text{HWHM}$, where HWHM is the half-width of half-maximum of the broadening Gaussian function in the frequency domain (HWHM$=120\ $cm$^{-1}$, as reported in the main text). Bottom: Absorption spectra computed as the Fourier transforms of the first 500 (blue, solid) or 1000 (red, dashed) steps of the damped autocorrelation function (blue, solid line in the top panel). For the given damping, 500 steps ($125\ $fs) of the autocorrelation function are sufficient to obtain a converged absorption spectrum, as demonstrated by the comparison of simulated spectra in the bottom panel.}
\label{fig:CorrelationConvergence}
\end{figure}
\newpage
\section{Shifts in frequency applied to the calculated spectra of phenol}

To facilitate the comparison between the calculated and experimental spectral lineshapes, we shift the simulated spectra along the frequency axis and scale their intensity so that all spectra match at the maximum of the experimental spectrum. To reproduce the exact position of the spectrum, one would need to predict accurately the vertical excitation energy, which is difficult even with fairly accurate electronic structure methods. As shown in Table~\ref{tab:vee}, values obtained at different levels of theory vary greatly. For example, the method used in this work, PBE0/6-311G(d,p), predicts the vertical excitation energy of $5.25\ $eV, multiconfigurational complete active space self-consistent field (CASSCF) method gives $4.85\ $eV, whereas the second-order perturbation theory restricted active space (RASPT2) method reduces the excitation energy to $4.47\ $eV. Based on the frequency shift that we employed to overlap the simulated thawed Gaussian and experimental spectra (see Table~\ref{tab:shifts}), we propose the value of $4.83\ $eV as the most accurate prediction of the vertical excitation energy. Note that this quantity cannot be deduced from the experimental spectrum. Some authors, e.g., in Ref.~\onlinecite{Nenov_Rivalta:2015}, incorrectly compared their results with the 0--0 transition energy ($36350\ $cm$^{-1} \approx 4.51\ $eV), which is typically much lower than the vertical energy gap at the Franck--Condon geometry; others, e.g., in Ref.~\onlinecite{Yang_Truhlar:2014}, compared their prediction with the transition energy of the maximum-intensity peak of the spectrum ($4.61\ $eV), which is not necessarily equal to the vertical excitation energy. Our prediction is, in fact, far from those values, emphasizing the importance of simulating full vibronic spectra for the accurate estimation of the vertical excitation energy.

By neglecting anharmonicity, Duschinsky rotation, or mode distortion, one inevitably modifies the estimated zero-point energy in the excited electronic state, which, in turn, affects the overall position of the absorption spectrum. Therefore, the spectra computed with different anharmonic and harmonic methods differ not only in shape but are also shifted differently. To compare fairly the spectral lineshapes, we translate each simulated spectrum with its own value of the frequency shift.

\begin{table}[t]
\caption{{Vertical excitation energies for the phenol S$_1 \leftarrow$ S$_0$ electronic transition evaluated at different levels of theory. Standard notation M$_1$/B$_1$//M$_2$/B$_2$ is used to denote the excitation energy evaluated with method M$_1$ and basis set B$_1$ at the ground-state minimum optimized with method M$_2$ and basis set B$_2$. For brevity, we do not list all values found in the literature and refer the reader to Refs.~\onlinecite{Gao_Freindorf:1996,Granucci_Tran-Thi:2000,Rogers_Hirst:2003,Zhang_Muchall:2006,Kim_Kang:2009,Yang_Truhlar:2014} for more computational results.}}
\label{tab:vee} \centering
\begin{ruledtabular}
\begin{tabular}{lc}
Electronic structure method & Vertical excitation energy (eV) \\ \hline
PBE0/6-311G(d, p)//PBE0/6-311G(d, p)						&	5.25\footnote[1]{This work.}	 \\
EOM-CCSD/6-311G(d, p)//PBE0/6-311G(d, p)					&	5.11\footnotemark[1] \\
CASSCF(10,10)/aug-cc-pVDZ//MP2\footnote[2]{Second-order M{\o}ller–Plesset perturbation theory.}/aug-cc-pVDZ				&	4.85\footnote[3]{Ref.~\onlinecite{Lan_Mahapatra:2005}} \\
MRCI\footnote[4]{Multireference configuration interaction.}/aug-cc-pVDZ//MP2/aug-cc-pVDZ	&	4.75\footnote[5]{Ref.~\onlinecite{Vieuxmaire_Domcke:2008}} \\
RASPT2(0,0/8,7/2,12)/ANO-L(432,21)-aug//CASSCF(8,7)/ANO	&	4.47\footnote[6]{Ref.~\onlinecite{Nenov_Rivalta:2015} See also other values reported therein.}
\end{tabular}
\end{ruledtabular}
\end{table}

\begin{table}[H]
\caption{Overall frequency shifts (in cm$^{-1}$) introduced into the calculated linear and two-dimensional spectra of phenol. Same shifts were used along both frequency axes in two-dimensional electronic spectra.}
\label{tab:shifts} \centering
\begin{ruledtabular}
\begin{tabular}{lc}
Method & Shift \\ \hline
On-the-fly ab initio thawed Gaussian approximation	&	-3380 \\
Harmonic approximation								&	-3440 \\
Uncoupled harmonic approximation						&	-3760 \\
Displaced harmonic oscillator model					&	-4740
\end{tabular}
\end{ruledtabular}
\end{table}

\bibliographystyle{aipnum4-2}
\bibliography{biblio48,additions_ThirdOrder}